\documentclass[11pt]{article}

\usepackage[final]{acl}

\usepackage{times}
\usepackage{latexsym}

\usepackage[T1]{fontenc}

\usepackage[utf8]{inputenc}

\usepackage{microtype}

\usepackage{inconsolata}
\usepackage{amsmath}
\usepackage{amssymb}
\usepackage{mathtools}
\usepackage{amsthm}
\usepackage{bbm}
\usepackage{enumitem}
\usepackage{graphicx}
\usepackage{booktabs} 

\usepackage{amsfonts} 
\usepackage{multirow}
\usepackage{arydshln}
\usepackage{colortbl}  
\usepackage{xcolor}    
\usepackage{tcolorbox}
\usepackage{tikz}

\usepackage[font=normalsize]{caption}

\usepackage{subcaption}
\usepackage{threeparttable}

\usepackage{longtable}
\title{R³A: Reinforced Reasoning for Relevance Assessment for RAG in User-Generated Content Platforms}

\author{
Xiaowei Yuan$^{1,2}$, Lei Jin$^{3}$, Haoxin Zhang$^{3}$, Ziyang Huang$^{1,2}$, \\\textbf{Yan Gao$^{3}$,} \textbf{Yi Wu$^{3}$,} \textbf{Yao Hu$^{3}$,} \textbf{Jun Zhao$^{1,2}$,} \textbf{Kang Liu}$^{1,2,*}$\\
$^1$The Key Laboratory of Cognition and Decision Intelligence for Complex Systems,\\
Institute of Automation, Chinese Academy of Sciences\\
$^2$School of Artificial Intelligence, University of Chinese Academy of Sciences\\
$^3$Xiaohongshu Inc.\\
}

\begin{document}
\maketitle
\begin{abstract}
Retrieval-augmented generation (RAG) plays a critical role in user-generated content (UGC) platforms, but its effectiveness critically depends on accurate query–document relevance assessment. Despite recent advances in applying large language models (LLMs) to relevance modeling, UGC platforms present unique challenges: 1) ambiguous user intent due to sparse user feedback in RAG scenarios, and 2) asymmetric relevance, where relevance is driven by localized answer-bearing content rather than global query–document similarity. To address these issues, we propose the Reinforced Reasoning model for Relevance Assessment (R³A), which decomposes relevance assessment into intent inference and evidence grounding. R³A leverages auxiliary high-clicked documents to infer latent query intent, and extracts verbatim evidence fragments to ground relevance decisions, reducing noise sensitivity and improving asymmetric relevance modeling. Experimental results demonstrate that R³A substantially outperforms strong baselines on offline benchmarks, while the distilled R³A-1.5B model achieves significant gains in large-scale online A/B testing, effectively balancing performance and practical deployability.
\end{abstract}

\section{Introduction}

Retrieval-augmented generation (RAG) systems have emerged as a critical paradigm in modern information retrieval, facilitating the generation of responses grounded in externally retrieved knowledge~\cite{ram-etal-2023-context,gao2024}. Within large-scale user-generated content (UGC) platforms—encompassing product reviews, travelogues, and lifestyle narratives—RAG serves a critical function in search architectures. By integrating retrieval and generation capabilities, these systems can efficiently search over billions of user-contributed documents and produce concise, informative responses tailored to user queries~\cite{li2025aisearchparadigm,zhang2025surveylargelanguagemodel}.

\begin{figure}[t]
  \centering
  \includegraphics[width=0.95\columnwidth]{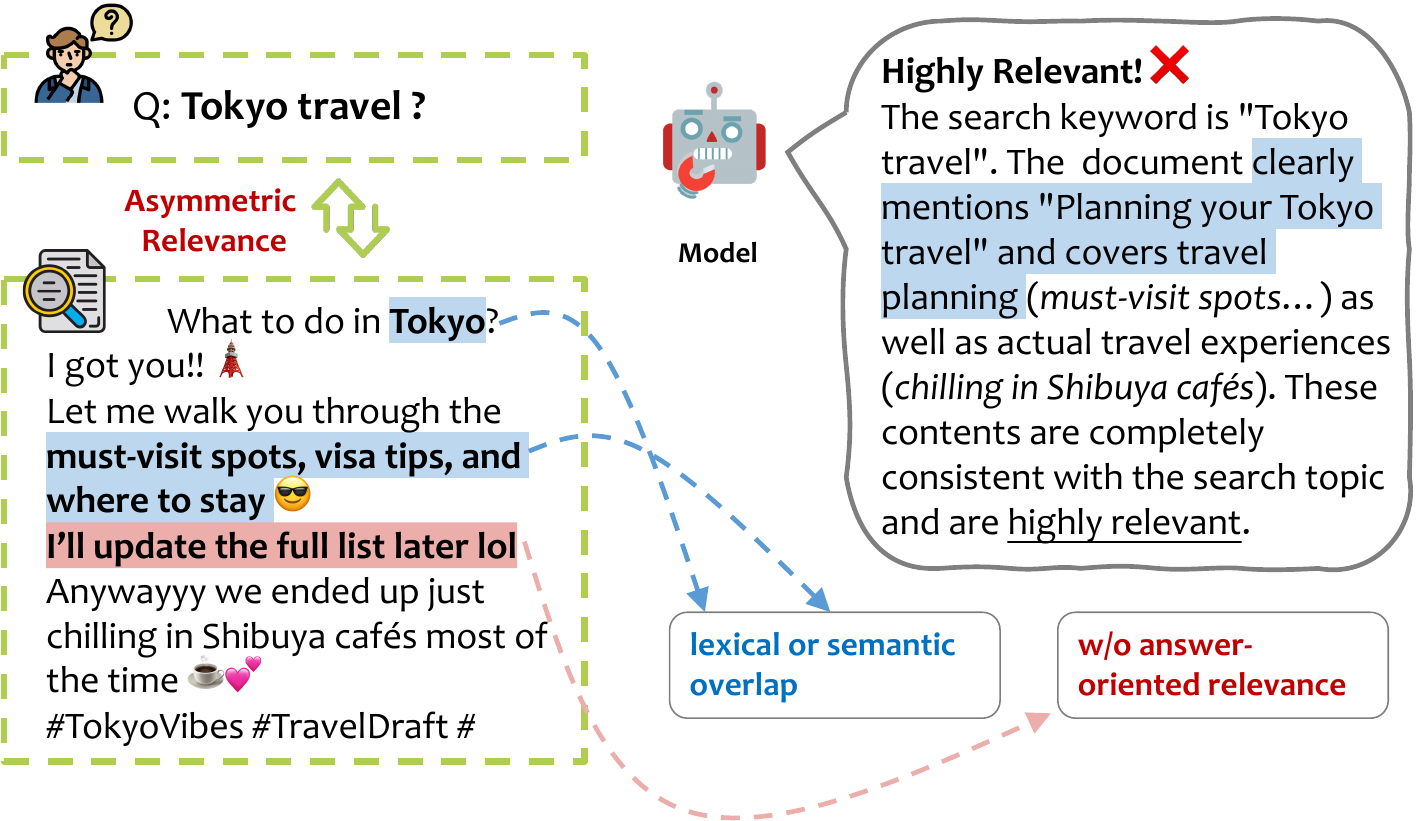}
  \caption{An illustrative example of erroneous relevance assessment for a query–document pair. The model is misled by superficial similarity in the noisy document, thereby failing to identify the absence of answer-bearing evidence required under asymmetric relevance.}
  \label{fig:intro}
\end{figure}

A critical component of the RAG system is the relevance assessment module. This module quantitatively evaluates the semantic relevance between user queries and retrieved documents~\cite{TomasBing}, ensuring that the generation is accurately grounded in the query-related context. Recent advances in large language models (LLMs) have opened up new possibilities for relevance modeling by enabling fine-grained understanding of semantics and intent~\cite{FaggioliDCDHHKK23,ZhengC00WZL0LXZ23}, relying on prompting~\cite{MacAvaneyS23,upadhya2} or supervised fine-tuning (SFT) methods~\cite{spm_meituan,Fitte}.

However, applying such models to UGC platforms introduces unique challenges. First, intent inference is constrained by the absence of click-through logs~\cite{SalemiZ24a,Feedback}, which conventional retrieval systems leverage to learn relevance–behavior mappings~\cite{jiang2024realbetteraligninglarge}. UGC-based RAG systems are typically supervised only at the response level, lacking the document-level interaction feedback necessary for effective \textbf{intent disambiguation}. In this context, intent represents the latent information need driving a query, which, on lifestyle-sharing UGC platforms, often transcends factual retrieval to encompass preferences for specific perspectives. As illustrated in Figure~\ref{fig:intro}, a query such as "\textit{Tokyo travel}" often implies a need for comprehensive guidance rather than isolated facts.
Second, RAG systems must reason over \textbf{asymmetric relevance}, where document usefulness depends on the presence of answer-bearing fragments rather than global query–document similarity. This issue is amplified in UGC due to its informal and noisy nature, which obscures semantic cues and leads to relevance misestimation. As shown in Figure~\ref{fig:intro}, the model is misled by superficial semantic overlap (e.g., “\textit{must-visit spots}”) despite the absence of substantive, answer-oriented content.

To address the above challenges, we propose a \textbf{R}einforced \textbf{R}easoning Model for \textbf{R}elevance \textbf{A}ssessment (R³A), which performs decomposed reasoning based on reinforcement learning (RL) algorithm.
We argue that generating high-quality relevance assessment in UGC scenarios requires strong reasoning capabilities to address the challenges of ambiguous query intent and asymmetric relevance with noisy document. To better capture the user intent, the model input is augmented with a set of auxiliary in-platform high-clicked documents retrieved using the same query. These additional documents provide contexts to help the model infer the user’s likely intent beyond the surface form of the query. 
Furthermore, the model is required to extract the most relevant answer fragment from the candidate document and determine relevance on the basis of this fragment.
By compelling the model to isolate and justify its assessment using the fragments, this approach not only mitigates the impact of noise but also facilitates the evaluation of asymmetric relevance, determining the presence or absence of answer-bearing content that satisfies the user’s information need.

Empirical results demonstrate that R³A consistently outperforms all baseline models in relevance assessment on both the industrial dataset NoteRel and the public benchmark T²Ranking~\cite{t2rank}. Furthermore, we distilled a compact R³A-1.5B model from the reinforced R³A-7B for online deployment. Notably, the distilled model not only exceeds the performance of its larger counterpart but also achieves significant gains over competing methods in large-scale online A/B testing.

The contributions of this paper are as follows:
\begin{itemize}[noitemsep]
    \item To tackle the unique challenges on the UGC platform, this paper proposes R³A method that performs decomposed reasoning over both ambiguous query and the asymmetric relevance with noisy document. It enhances reasoning for query disambiguation and evidence localization. 
    \item On offline experiments, R³A consistently surpasses strong baselines. The distilled R³A-1.5B model outperforms its larger counterpart in online A/B testing, demonstrating the practical effectiveness of proposed R³A method.
\end{itemize}

\section{Decomposed Reasoning for Relevance Assessment}
This paper proposes the \textbf{R}einforced \textbf{R}easoning Model for \textbf{R}elevance \textbf{A}ssessment (R³A) method for UGC platforms, which enhances the reasoning capabilities of relevance modeling.

\begin{figure}[t]
    \centering
  \includegraphics[scale=0.24]{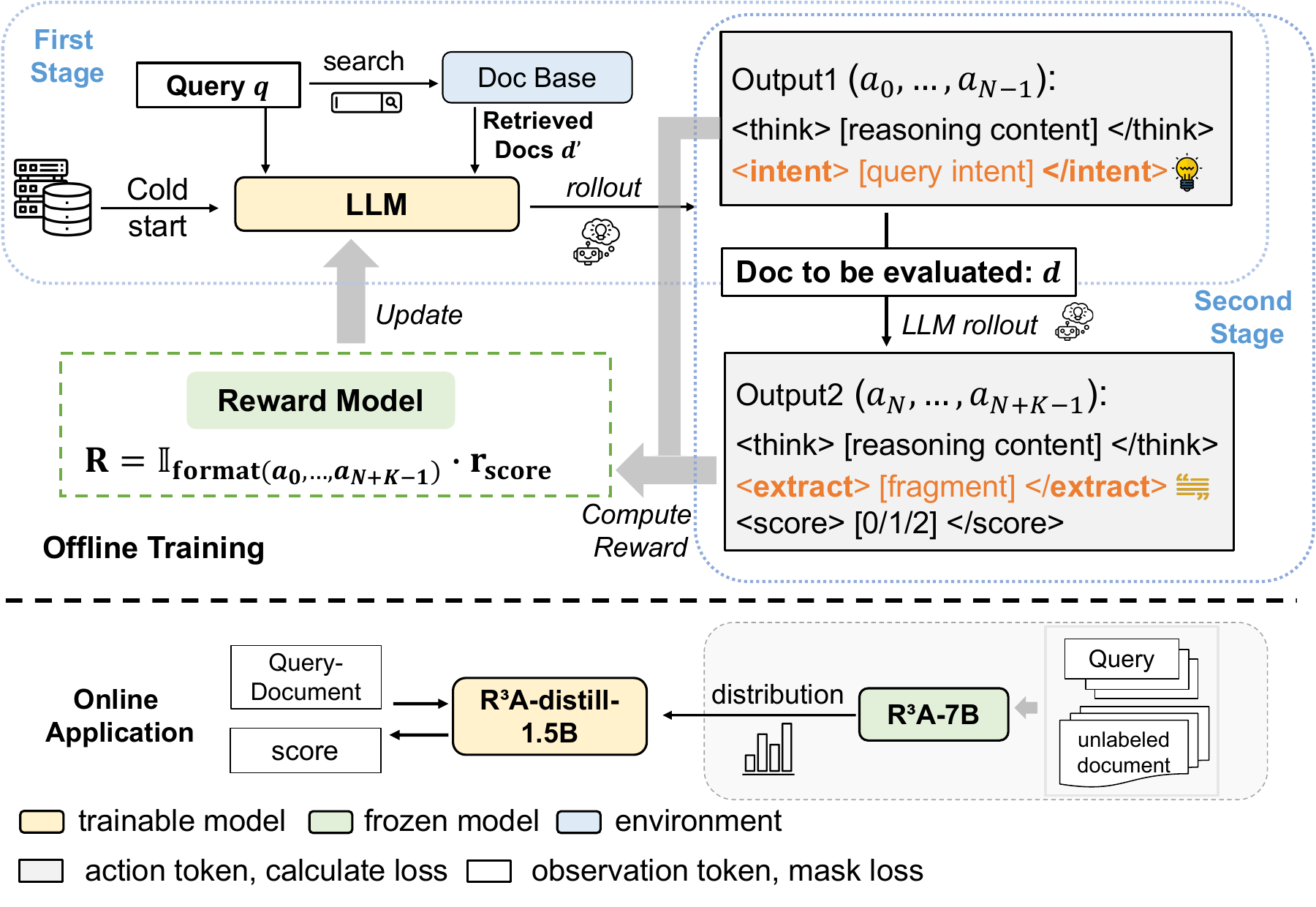}
  \caption{The overview of the R³A framework.}
  \label{fig:method}
\end{figure}

The overall framework of R³A is illustrated in Figure~\ref{fig:method}. After a cold-start initialization, the RL training procedure involves a two-stage interaction between the model and the environment (in-platform documents).
\textit{In the first stage}, a set of auxiliary in-platform documents $d' (\leq 4)$ is provided alongside the user query $q$ to support the model in inferring latent query intent. 
\textit{In the second stage}, the model extracts answer-bearing fragment from $d$ to model the asymmetric relevance and ground its assessment in semantically aligned content.

For online deployment, we distill label logits from a reinforcement-trained large R³A model into a smaller model, which directly outputs relevance scores without generating reasoning traces.

\subsection{Cold Start}
Following prior work~\cite{deepseek,wei2025cold,chen2025cold}, we first perform a cold-start phase using a 50k unlabeled dataset (see Section~\ref{sec:datasets}). This cold-start training on the relevance assessment task is designed to instill structured reasoning behavior and the desired output format, thereby improving training stability before RL. The structured chain-of-thought reasoning outputs are generated by DeepSeek-R1~\cite{deepseek}\footnote{Detailed prompt is provided in Appendix~\ref{appendix:umbrela_prompt}.}. 

\subsection{Decomposed Reasoning with RL} After the cold-start initialization, R³A is trained via the GRPO~\cite{grpo} algorithm. By estimating advantages through group-relative rewards instead of a dedicated critic model (as in PPO~\cite{ppo}), GRPO facilitates more efficient large-scale training. We use GRPO primarily as a high-performance realization tool to showcase the improvements brought by our reasoning framework. 

GRPO performs multiple rollouts per input and calculate the relative reward $r$ within the group as the advantage $A$. It optimizes the following objective:

\vspace{-2.0em}
\begin{small}
\begin{equation}
    \mathcal{J}(\theta)=\mathbb{E}_{(q,d) \sim P(Q,D), \{\tau_i\}^{G}_{i=1} \sim \pi_{\theta_{old}}} \bigg[\frac{1}{|G|}\sum_{i=1}^{|G|}  \nonumber
\end{equation}
\end{small}
\vspace{-2.5em}
\begin{small}
\begin{align}
     \Big(\frac{1}{|\tau_i|}\sum_{t=1}^{|\tau_i|}  \min \Big( r_\theta A_i, \text{clip} \big( r_\theta, 1 - \epsilon, 1 + \epsilon \big)  A_i \Big)\bigg], \nonumber
\end{align}
\end{small}
\vspace{-1.0em}
\begin{equation} \label{eq:grpo}
r_{\theta}=\frac{\pi_\theta(a_{i,t} | x, d, a_{i,<t})}{\pi_{\theta_{old}}(a_{i,t} | x, d, a_{i,<t})} 
\end{equation}
where $q$ and $d$ denote the query and associated document sampled from the training distribution.
Given an input $(q, d)$ pair\footnote{The retrieved document $d'$, based on the input query and document $d$, is omitted here for brevity.}, a group $G$ of trajectories ${\tau_i}$ is generated using the old policy $\pi_{\theta_{\text{old}}}$. Each trajectory $\tau_i$ comprises a sequence of actions $a_i$, representing the output of model reasoning. $\epsilon$ is the clipping ratio. 

The term $A_i = \frac{r_i - \mu_r}{\sigma_r}$ represents the standardized advantage of trajectory $\tau_i$, where $r_i$ is the reward assigned to the trajectory, and $\mu_r$ and $\sigma_r$ are the mean and standard deviation of the rewards within group $G$. 

\paragraph{Reasoning on Query Intent.} 
In the initial stage, the model interacts with a set of auxiliary, highly-clicked documents $d' (|d'|\leq 4)$ retrieved using the original query $q$ from the platform. These documents supply additional contextual signals that enable the model to infer the user’s latent intent beyond the query’s surface form.

So the trajectory in the first stage can be represented as: $\tau_{i_1} = (q,d',a_{i,0},a_{i,1}, ...,a_{i,N-1})$, where $N$ denotes the number of action tokens. The action is structured token sequence using special tags, as pioneered by recent works~\cite{format1,format2,format3}. Each action must contain a reasoning step (\textsc{<think>}) and an intent step (\textsc{<intent>}). The detailed system template is shown in Appendix~\ref{appendix:system_prompt_1}.

\paragraph{Reasoning with Asymmetric Relevance.} 

As asymmetric relevance depends on the presence of answer-bearing segments rather than overall query–document similarity, the model is tasked with extracting verbatim, query-relevant fragments from candidate documents or returning \textsc{None} if no match exists. This formulation compels the model to ground its assessments in specific textual evidence, thereby ensuring directional, answer-centric judgments while mitigating the impact of noise.

In the extraction process, we enforce correctness strictly at the structural level—ensuring verbatim consistency with the document fragments—owing to the absence of ground-truth evidence annotations. The semantic accuracy and answer-oriented alignment of these extractions are intended to emerge through reinforcement-based reasoning training, which leverages available relevance labels to guide the model's optimization.

Thus, the complete trajectory in a rollout (two-stage interaction) can be represented as $\tau_i = (\tau_{i_1}, d, a_{N,i},$ $ a_{N+1,i}, \ldots, a_{N+K-1,i}, r)$, where $K$ denotes the number of tokens and $r$ denotes the reward to be calculated. Each action must contain a reasoning step (\textsc{<think>}), an extraction step (\textsc{<extract>}) and a final answer (\textsc{<score>}). The detailed system template is shown in Appendix~\ref{appendix:system_prompt_2}.

\paragraph{Reward Function.}
We design the rule-based reward function such that a reward is granted if and only if the LLM-generated output fully conforms to all specified reasoning and answer formats, as well as the extraction consistency requirement. The total reward $R$ is defined as:

\vspace{-1.0em}
\begin{small}
\begin{align}
    R = \mathbb{I}_{\text{format}} \cdot r_{score}, \quad
    &r_{\text{score}} =
\begin{cases}
1 & \text{$s_{pred}$ == $s_{gold}$} \\
\lambda & \text{|$s_{pred}$-$s_{gold}|=1$} \\
0 & \text{|$s_{pred}$-$s_{gold}|=2$} \\
\end{cases}
\end{align}
\end{small}
where $\mathbb{I}_{\text{format}}$ is an indicator function that equals 1 if the trajectory format is correct, and 0 otherwise. The $r_{\text{score}}$ measures the correctness of the model's prediction $s_{\text{pred}}$ compared to the gold score $s_{\text{gold}}$. The hyperparameter $\lambda \in [0, 1)$ is introduced to impose a soft penalty. The value of $\lambda$ balances the trade-off between strict correctness and leniency in reward shaping. In Exp.~\ref{sec:ablation}, we investigate the impact of varying $\lambda$ on model performance.

\subsection{Distillation} 
For online deployment, although RL with explicit reasoning can substantially improve performance, it also incurs prohibitive inference costs due to the generation of long reasoning traces. To reconcile these objectives, we deploy a distillation scheme that preserves the teacher’s performance capabilities while eliminating the runtime overhead of trace generation. Concretely, we use a large reinforced R³A teacher (R³A-7B) to distill into a compact 1.5B student model on 50k samples. The distilled 1.5B model takes the query-document pair as input and directly outputs the relevance score. Unlike the teacher model, it does not generate reasoning traces, which minimizes online latency.

Formally, let $p_S(\cdot \mid x; \theta_S)$ denote the student’s predictive distribution obtained by applying a softmax to its logits, and let $p_T(\cdot \mid x; \theta_T)$ be the final score distribution from the teacher model. The training objective is:
\begin{equation}
\small
    \mathcal{L} = -\mathbb{E}_{x \sim \mathcal{D}} \left[ \sum_{k} p_T(k \mid x; \theta_T) \log p_S(k \mid x; \theta_S) \right] 
\end{equation}

\section{Experiment}
In this section, we conduct both offline and online experiments to evaluate the performance of the proposed R³A method.
\subsection{Datasets} \label{sec:datasets}

We evaluate relevance assessment performance on both an internal industrial dataset, \textbf{NoteRel}, and a public benchmark, \textbf{T\textsuperscript{2}Ranking}. Summary statistics for the two datasets are reported in Table~\ref{tab:dataset}.

\paragraph{NoteRel.} 
To assess relevance in a practical setting, we constructed NoteRel, a dataset derived from our online RAG system deployed on Xiaohongshu.

We extracted real-world user queries and their corresponding retrieved documents from online system logs. To generate high-quality training samples, we utilized document citation signals as a preliminary filter. For each query and candidate documents, the response generator performed $M$ forward passes. A document was classified as a high-confidence positive sample if it was cited in at least $N$ generated responses; otherwise, it was treated as a challenging hard negative. In our implementation, we set $M=5$ and $N=2$. To further account for distributional noise, we supplemented the dataset with random negatives sampled from the global corpus.

To establish "gold" standard labels, we collected 7k samples for professional human annotation across three categories: \textbf{0-Irrelevant, 1-Partially Relevant, and 2-Highly Relevant}. A detailed description of the dataset and data preparation procedures is provided in Appendix~\ref{appendix:dataset}. Additionally, 50k unlabeled documents were collected to support cold-start training and distillation phases.


\paragraph{T²Ranking.} 
To ensure the generalizability of our framework, we also utilized T²Ranking, a large-scale retrieval benchmark comprising over 300k real-world queries and web passages. For our experiments, we sampled 3k records from the test set. 
T²Ranking provides fine-grained relevance labels on a 4-level scale (0–3). To maintain consistency with the NoteRel schema, we aligned the 2-level and 3-level labels in T²Ranking with the Highly Relevant (2) class in our industrial dataset.

\begin{table}
  \centering
  \small
  \begin{tabular}{lcccc}
    \hline
    \multirow{2}{*}{\textbf{Datasets}} & \multicolumn{2}{c}{\textbf{NoteRel}} & \multicolumn{2}{c}{\textbf{T²Ranking}}\\
     & \textbf{Training} & \textbf{Test} & \textbf{Training} & \textbf{Test}\\
    \hline
    \#0      & 2002 & 300  & \verb|\| & 1000       \\
    \#1      & 1858 & 300  & \verb|\| & 1000       \\
    \#2      & 1999 & 300  & \verb|\| & 1000       \\
    \hline
  \end{tabular}
  \caption{Summary statistics of datasets.}
  \label{tab:dataset}
\end{table}

\subsection{Settings}
\paragraph{Baselines.} 
To demonstrate the effectiveness of R³A, we compare it with the following baselines\footnote{All methods are conducted in a zero-shot manner.}: (1) \textbf{Prompting (UMbrela)}~\cite{upadhya2}: It is a prompting-based method that provides a step-by-step guide to structure the relevance labeling task, thereby facilitating more nuanced reasoning by the LLM. In our experiment, we reproduce the UMbrela method using several LLMs, including QwQ~\cite{qwen2024qwq}, DeepSeek-R1~\cite{deepseek}, GPT-4o~\cite{gpt4o}, among others. (2) \textbf{SFT}~\cite{Fitte}: This method introduces a framework that directly fine-tunes the model using relevance label. To ensure a fair comparison, we also pre-train the model using cold-start data before fine-tuning. (3) \textbf{R1-Zero/R1}~\cite{deepseek}: This method employs rule-based GRPO to encourage the model to engage in explicit reasoning during relevance assessment. The model's output format is constrained to the \textsc{<think>} and \textsc{<score>} tags. The R1-Zero refers to RL initiated without the cold-start strategy. (4) \textbf{Distillation}: The distilled model is trained on a 1.5B backbone using the same SFT approach. It relies solely on score labels produced by the R³A-7B model on 50k cold-start data. We deploy this version in our online system for better inference speed and overall throughput.

\begin{table}[t]
\centering
\resizebox{0.48\textwidth}{!}{%
\begin{tabular}{@{}lcccccc|cccccc@{}}
\toprule
& \multicolumn{6}{c}{\textbf{NoteRel}} & \multicolumn{6}{c}{\textbf{T²Ranking}} \\
\cmidrule(lr){2-7} \cmidrule(lr){8-13}
\multirow{2}{*}{\textbf{Method}} & \multicolumn{3}{c}{\textbf{F1-Score}} & \multicolumn{2}{c}{\textbf{AUC}} & \multirow{2}{*}{\textbf{Acc.}}  & \multicolumn{3}{c}{\textbf{F1-Score}} & \multicolumn{2}{c}{\textbf{AUC}} & \multirow{2}{*}{\textbf{Acc.}} \\
\cmidrule(lr){2-4} \cmidrule(lr){5-6} \cmidrule(lr){8-10} \cmidrule(lr){11-12} 
& 0 & 1 & 2 & 0/12& 01/2& & 0 & 1 & 2 & 0/12& 01/2  \\
\midrule
\multicolumn{7}{@{}l}{\textbf{Prompting}} \\
\multicolumn{7}{@{}l}{\textit{UMbrela w/o parameter update}} \\
Qwen2.5-7B & 48.1 & 46.6 & 54.6 & 65.1 & 66.3 & 49.6 & 47.7 & 30.1 & 60.0 & 65.2 & 65.9 & 48.9 \\
QwQ-32B & 62.3 & 43.6 & 63.5 & 70.3 & 69.4 & 55.9 & 48.2 & 33.8 & \textbf{67.1} & 65.5 & 66.9 & 54.0 \\
DeepSeek-V3 & 56.6 & 46.3  & 60.4 & 69.4 & 69.8 & 54.1 & 49.8 & 32.7 & 65.8 & 66.5 & 67.6 & 52.3\\
DeepSeek-R1 & 61.9 & 42.8 & 63.0 & 72.1 & 71.8 & 56.0 & 46.5 & 31.9 & 66.9 & 64.3 & 63.2 & 53.6 \\
GPT-4o  & 63.3 & 53.8 & 59.6 & 73.0 & 69.8 & 58.2 & 50.9 & 33.4 & 66.2 & 66.3 & 67.1 & \textbf{54.2}\\
\midrule\midrule 
\multicolumn{7}{@{}l}{\textbf{Supervised Fine-Tuning}} \\
\multicolumn{7}{@{}l}{\textit{w/ parameter update}} \\
Qwen2.5-1.5B & 68.4 & 51.8 & 57.7 & 76.3 & 68.5 & 59.0 & 48.3 & 36.0 & 50.4 & 67.6 & 62.9 & 46.3\\
+ pretrained & 70.1 & 55.3 & 56.0 & 77.3 & 68.2 & 60.0 & 50.7 & 33.3 & 58.2 & 69.1 & 65.4 & 48.3\\
Qwen2.5-7B & 71.0 & 55.6 & 52.1 & 78.2 & 66.8 & 60.3 & 49.7 & 32.2 & 62.8 & 66.4 & 67.5 & 50.7\\
+ pretrained & 69.9 & 54.9 & 53.9 & 77.0 & 67.6 & 61.4 & 51.2 & 31.6 & 65.3 & 66.9 & 67.5 & 51.4\\
\midrule\midrule 
\multicolumn{7}{@{}l}{\textbf{Reinforcement Learning}} \\
\multicolumn{7}{@{}l}{\textit{w/ parameter update on Qwen2.5-1.5B}} \\
R1-Zero & 66.7 & 45.3 & 58.5 & 75.1 & 68.2 & 56.8 & 47.9 & 26.8 & 62.3 & 65.0 & 64.1 & 46.2 \\
R1 & 70.7 & 46.9 & 62.0 & 78.3 & 69.0 & 59.7 & 51.1 & 25.4 & 63.6 & 65.6 & 64.8 & 47.7\\
\rowcolor{blue!5} R³A-Zero & 71.3 & 49.8 & 60.5 & 79.4 & 68.3 & 60.4 & 48.0 & 24.8 & 61.8 & 64.4 & 65.0 & 47.9 \\
\rowcolor{blue!5} R³A & 72.0 & 51.5 & 62.9 & 80.8 & 69.6 & 61.7 & 49.8 & 30.3 & 62.7 & 66.1 & 66.4 & 49.0 \\
\midrule 
\multicolumn{7}{@{}l}{\textit{w/ parameter update on Qwen2.5-7B}} \\
R1-Zero & 67.5 & 47.2 & 62.5 & 75.6 & 71.7 & 59.3 & 48.0 & 24.8 & 61.8 & 64.4 & 65.0 & 49.9 \\
R1 & 74.4 &  49.6 & 63.7  & 81.2 & 72.7 & 63.2 & 50.8 & 27.3 & 57.7 & 66.4 & 65.6 & 51.4\\
\rowcolor{blue!5} R³A-Zero & 75.8 & 51.7 & 63.3 & 82.4 & 72.5 & 63.6 & 50.8 & 33.2 & 63.5 & 67.3 & 64.8 & 51.2 \\
\rowcolor{blue!5} R³A & \textbf{77.1} & \textbf{56.0} & \textbf{64.2} & \textbf{83.1} & \textbf{73.3} & \textbf{65.2} & \textbf{52.1} & \textbf{34.3} & 66.6 & \textbf{67.5} & \textbf{67.8} & 53.9 \\
\midrule\midrule 
\multicolumn{7}{@{}l}{\textbf{Distilling (Online Serving)}} \\
\rowcolor{blue!5} R³A-Distill-1.5B & 71.4 & 55.9 & 60.3 & 78.3 & 70.5 & 62.0 & 50.4 & 29.8 & 64.9 & 66.3 & 67.2 & 51.3\\
\bottomrule
\end{tabular}
} 
\caption{Overall performance on the test set of NoteRel and T²Ranking datasets. 
The labels 0, 1, and 2 indicate "Irrelevant", "Partially Relevant", and "Highly Relevant", respectively. 
}
\label{tab:main_table}
\end{table}
\begin{figure}[t] 
    \centering
    \begin{subfigure}{0.23\textwidth}
        \includegraphics[width=\textwidth]{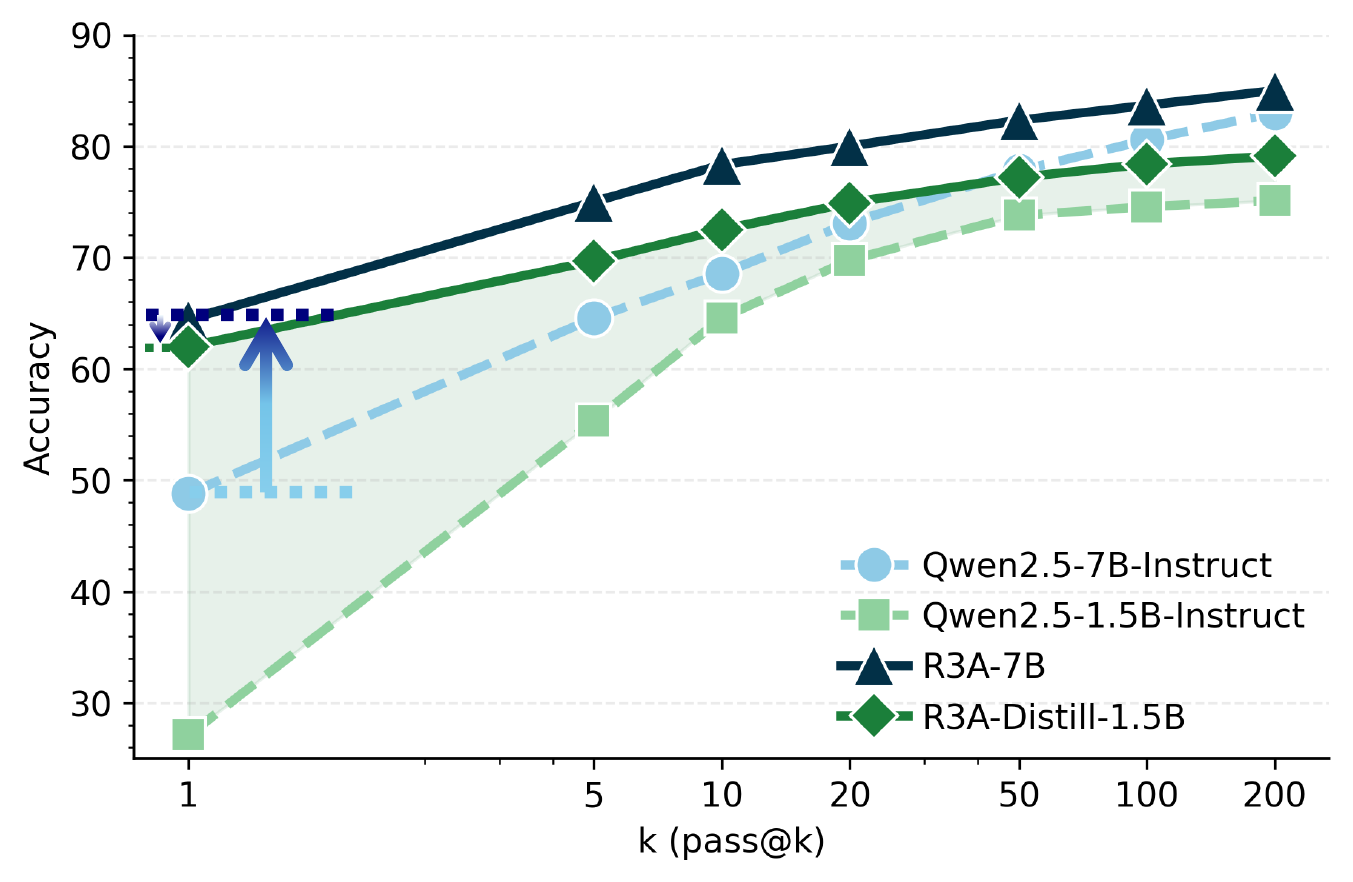}
        \caption{Pass@k performance comparison across different models}
        \label{fig:pass}
    \end{subfigure}
\hfill
    \begin{subfigure}{0.23\textwidth}
        \includegraphics[width=\textwidth]{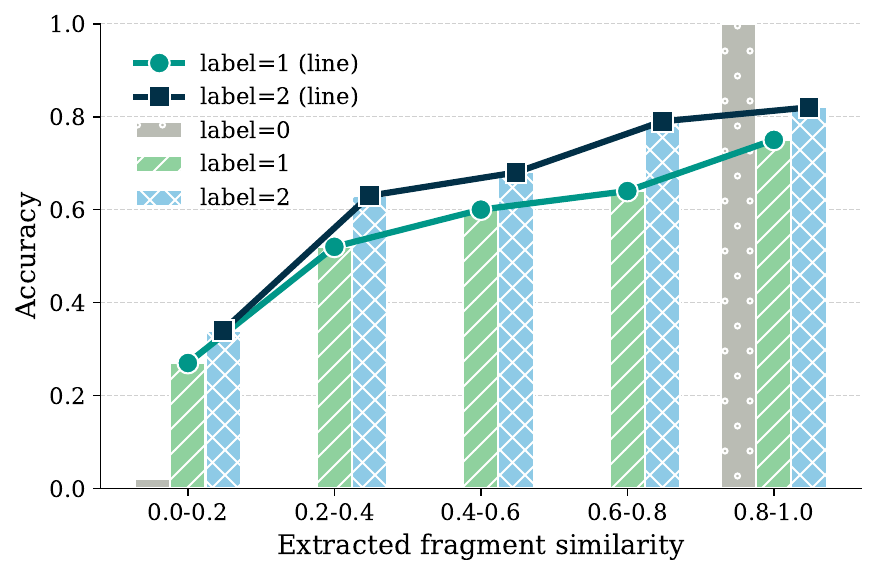}
        \caption{Correlation between extracted fragment similarity and accuracy}
        \label{fig:extract}
    \end{subfigure}
    \caption{Performance analysis of R³A model.}
    \label{fig:performance}
\end{figure}

\paragraph{Models.} For SFT and RL-based methods, we explore LLMs using instruction-tuned Qwen2.5 models~\cite{Qwen2.5} ranging from 1.5B to 7B parameters. All models are trained on the training split of the NoteRel dataset, with detailed training configurations provided in Appendix~\ref{appendix:training} and training logs presented in Appendix~\ref{appendix:log_fig}.

\paragraph{Metrics.} 
We use F1 score (macro-averaged), Accuracy, and one-vs-rest AUC metrics (AUC\textsubscript{0/12} and AUC\textsubscript{01/2}) following previous works~\cite{auc,chen2024towards}.

\begin{table} [t]
\centering
\tiny
\resizebox{0.48\textwidth}{!}{
\begin{tabular}{cccccc}
\hline
\textbf{Time / 8 A100 GPU} & \multicolumn{5}{c}{\textbf{Number of Evaluated Documents}} \\
\cline{2-6}
\textbf{(in seconds)} & \textbf{1} & \textbf{10} & \textbf{100} & \textbf{1,000} & \textbf{10,000}\\
\hline
R³A-7B &\textcolor{blue}{0.6385} & 
\textcolor{blue}{1.1718} & 
\textcolor{blue}{3.3774} & 
\textcolor{blue}{43.7740} & 
\textcolor{blue}{493.7861}\\
R³A-Distill-1.5B &\textcolor{green!60!black}{0.1396} & 
\textcolor{green!60!black}{0.1717} & 
\textcolor{green!60!black}{0.3332} & 
\textcolor{green!60!black}{2.3673} & 
\textcolor{green!60!black}{10.9421}\\
\hline
\end{tabular}
}
\caption{Inference latency comparison of R³A-7B and R³A-Distill-1.5B models.}
\label{tab:time}
\end{table}

\subsection{Overall Performance} \label{sec:main_results}
As presented in Table \ref{tab:main_table}, our method R³A consistently achieves superior performance over all baseline approaches on the NoteRel test set. Moreover, on the out-of-distribution benchmark T²Ranking, R³A achieves performance comparable to GPT-4o, indicating that the proposed approach generalizes effectively to datasets from other commercial domains. We present several cases in Appendix~\ref{appendix:case_study}.

Notably, the R³A-Zero models, trained without cold-start initialization, already surpass both R1-Zero and SFT baselines, underscoring the efficacy of R³A in exploiting reasoning signals to improve performance even under constrained initialization settings. When full RL training is applied, R³A further extends its advantage, yielding state-of-the-art results across most evaluation metrics. 


\paragraph{Superior Distillation.} 
Remarkably, the distilled model R³A-Distill-1.5B not only retains the performance gains of its 1.5B RL-trained counterpart, but also outperforms the larger 7B SFT model. These results indicate that the knowledge distilled from R³A effectively preserves essential relevance assessment capabilities in a smaller model.

As shown in Table~\ref{tab:time}, we compare inference time before and after model distillation on NoteRel dataset. The results indicate that the R³A-7B model’s inference latency increases almost linearly with the number of evaluated documents. In contrast, distillation substantially reduces inference cost by eliminating the need to output intermediate reasoning steps, thereby mitigating latency and improving efficiency.


\paragraph{Performance on Pass@k}
Figure~\ref{fig:pass} shows that while base instruction-tuned models improve with increasing $k$, their performance remains substantially below that of R³A-trained models, particularly at low $k$ (e.g., pass@1 and pass@5), where robust reasoning is critical. Consistent with prior findings~\cite{deng2025trialanderrorimprovementsystematicanalysis}, RL yields pronounced gains in low-sampling regimes, enabling R³A-7B to consistently outperform all baselines across pass@$k$. Notably, the distilled R³A-Distill-1.5B model closely approaches its 7B teacher and achieves over 10\% absolute improvement over Qwen2.5-1.5B-Instruct at pass@1, demonstrating effective knowledge transfer and strong deployability under efficiency constraints.
\paragraph{Role of Extraction in Reasoning}
Figure~\ref{fig:extract} analyzes the relationship between the similarity of reasoning fragments extracted by R³A-7B and those generated by GPT-4o under identical prompts, and its impact on relevance prediction accuracy. The results show a strong positive correlation, with near-optimal performance at high similarity levels (0.8–1.0) and sharp degradation at low similarity (0.0–0.2), particularly for partially relevant cases (score = 1). These findings highlight fragment extraction as a central mechanism in R³A, enabling accurate asymmetric relevance assessment while reducing the influence of document noise.

\begin{table}[t!]
\centering
\begin{threeparttable}
\resizebox{0.45\textwidth}{!}{%
\begin{tabular}{@{}lcccccccccccc@{}}
\toprule
\multirow{2}{*}{\textbf{Method}} & \multicolumn{3}{c}{\textbf{F1-Score}} & \multicolumn{2}{c}{\textbf{AUC}} & \multirow{2}{*}{\textbf{Accuracy}}   \\
\cmidrule(lr){2-4} \cmidrule(lr){5-6} 
& 0 & 1 & 2 & 0/12& 01/2&  \\
\midrule
\textbf{R³A-7B ($\lambda=0$)}  & \textbf{77.1} & \textbf{56.0} & \textbf{64.2} & \textbf{83.1} & \textbf{73.3} & \textbf{65.2} \\
\midrule\midrule
\multicolumn{7}{@{}l}{\textit{Format Variants}} \\
\quad w/o \texttt{<intent>}     & 76.0 & 55.9 & 63.7 & 82.6 & 72.8 & 64.7 \\
\quad w/o \texttt{<extraction>}  & 74.9 & 52.7 & 63.2 & 81.1 & 72.4 &63.9 \\
\midrule\midrule
\multicolumn{7}{@{}l}{\textit{Input Variant$^*$}} \\
\quad w/o \texttt{retrieval}  &  75.7 & 54.4 & 63.9 & 82.4 & 72.9 & 64.6\\
\quad w/ \texttt{half noisy docs}  & 76.3 & 55.5 & 63.3 & 82.7 & 72.6 & 64.8\\
\quad w/ \texttt{all noisy docs}  & 75.4 &54.2 & 63.3 &81.9 & 71.8 & 64.3 \\
\midrule\midrule
\multicolumn{7}{@{}l}{\textit{Reward Variants}} \\
\quad $\lambda=0.5$ & 75.3 & 52.5 & 59.7 & 82.0 & 69.9 & 62.0\\
\quad $\lambda=0.2$ & 74.9 & 53.6 & 62.9 & 81.9 & 72.3 & 62.5\\
\quad $\lambda=0.1$ & 72.6 & 48.4 & 60.3 & 80.2 & 70.8 & 61.4 \\
\midrule\midrule
\multicolumn{7}{@{}l}{\textit{Interaction Stage Variant}} \\
\phantom{0} single stage $\dagger$ & 73.8 & 49.1 & 58.3 & 80.7 & 68.5 & 60.6 \\
\bottomrule
\end{tabular}
} 
\begin{tablenotes}
\begin{minipage}{0.45\textwidth}
\tiny
\item[*] "Without retrieval" denotes the removal of retrieved in-platform documents. "half/all noisy docs" denote that half or all of the retrieved documents are randomly replaced with irrelevant ones.
\item[$\dagger$] Both the retrieved documents and the document to be evaluated are input into the model, which is required to analyze the intent, extract the fragment and assess the relevance.
\end{minipage}
\end{tablenotes}
\normalsize
\end{threeparttable}
\caption{Ablation study on the R³A method on NoteRel dataset.}
\label{tab:r3a_ablation}
\end{table}

\subsection{Ablation Study} \label{sec:ablation}
Results in Table~\ref{tab:r3a_ablation} validate each component's contribution. \textit{Format Variants} reveal that omitting extraction reasoning causes the most significant degradation, emphasizing the necessity of document-based grounding. Accurate \textit{retrieval in input} enhances intent inference by supplying informative contextual cues. Regarding \textit{Reward Variants}, $\lambda=0$ is optimal for maintaining strict class differentiation. Lastly, collapsing the Two-stage Interaction into a single stage reduces accuracy to 60.6\%, suggesting that excessive context length diminishes the model's focus on the document to be evaluated.


\begin{figure}[t]
    \centering
  \includegraphics[scale=0.24]{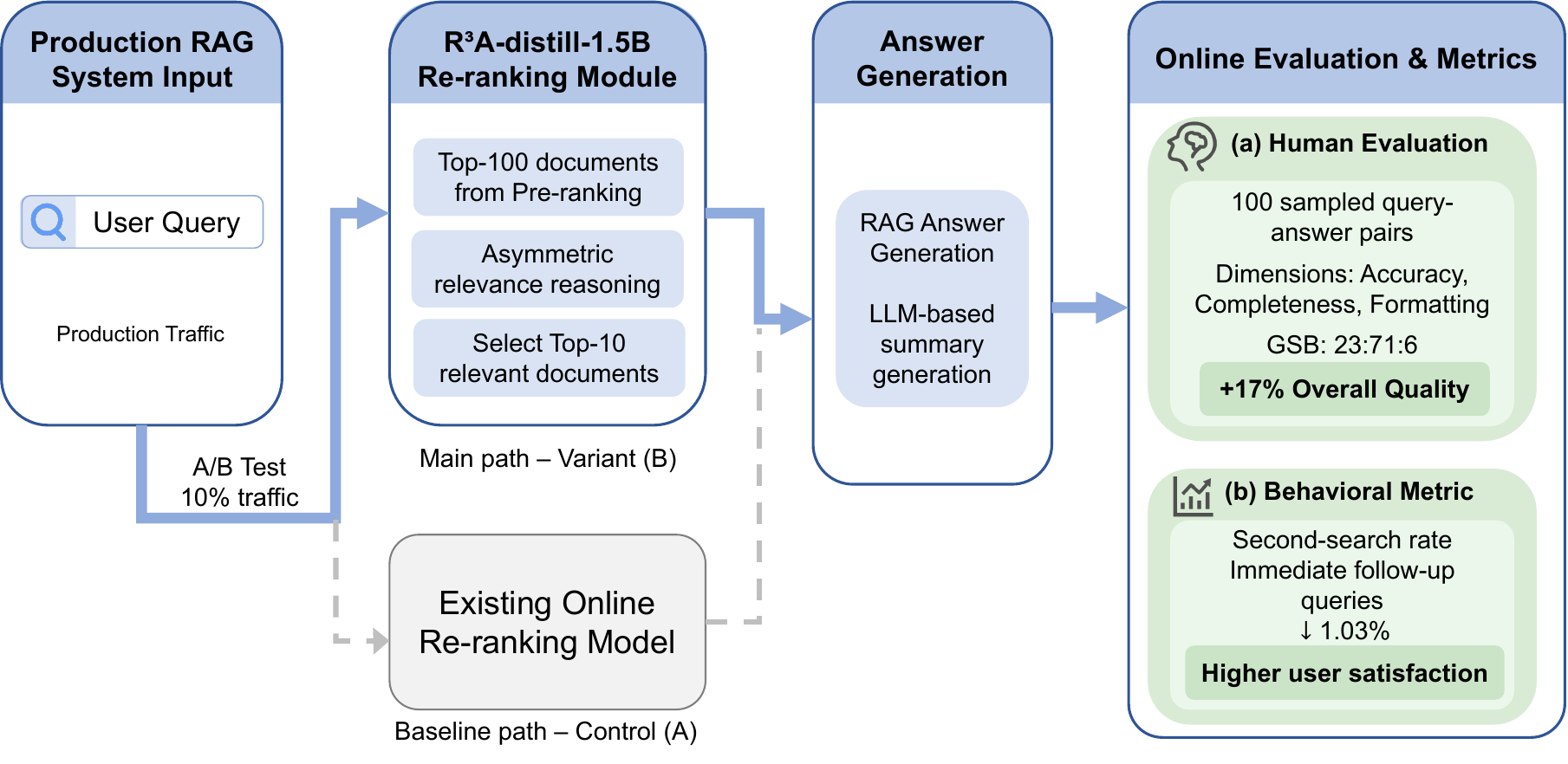}
  \caption{Overview of the online deployment and evaluation pipeline of R³A-Distill-1.5B.}
  \label{fig:online}
\end{figure}
\subsection{Online Performance}
As illustrated in Figure~\ref{fig:online}, we deploy \textit{R³A-Distill-1.5B} as the re-ranking module in the RAG system of Xiaohongshu and evaluate it via a one-week online A/B test on 10\% of live traffic. The model re-ranks the top 100 pre-retrieved documents and selects the top 10 for answer generation.

We first perform a human evaluation using 100 randomly sampled query–answer pairs from the system logs. 
Compared with our existing online model~\cite{Fitte}, the distilled model achieves a Good:Same:Bad (GSB) distribution of $23:71:6$, corresponding to a 17\% improvement in overall answer quality. Moreover, using second-search rate as a proxy for user satisfaction, the distilled model achieves a 1.03\% reduction, indicating improved effectiveness and reduced need for query reformulation.

By eliminating runtime reasoning trace generation, the distilled model achieves an average response time (RT) of approximately 150ms. Each single GPU in the production cluster handles a throughput of 200 QPS (Queries Per Second). To support the massive live traffic of the platform, we have deployed the distilled model on a cluster of 1,000 GPUs.

\section{Conclusion}
This paper proposes the R³A method, a novel decomposed reasoning framework tailored for relevance assessment for RAG system in UGC scenarios. 
Empirically, R³A exhibits strong capabilities in relevance assessment task in UGC scenarios.

\section*{Limitations}

Despite the effectiveness of R³A in enhancing relevance assessment for RAG systems within UGC scenarios, several limitations remain. R³A is primarily trained on an industry-specific UGC dataset. As demonstrated in Exp.\ref{sec:main_results}, our method consistently outperforms all baseline approaches on the NoteRel test set. On the out-of-distribution benchmark T²Ranking, the performance gap of R³A is notably smaller than that observed on NoteRel. During in-domain training, the model learns to reason about user query intent within UGC communities. However, UGC platforms and general-purpose search engines exhibit distinct contextual characteristics. When the model applies reasoning patterns learned from UGC data to assess relevance between queries and web passages in non-UGC settings, the contextual mismatch limits the magnitude of performance gains. 
Nevertheless, the proposed approach demonstrates reasonable generalization to datasets from other commercial domains and remains computationally efficient and well-suited for RAG applications in UGC scenarios.

\bibliography{custom}
\clearpage
\appendix

\section{Related Work}
Relevance modeling evaluates the extent to which a document satisfies a user query. Traditional human-annotated approaches are costly and prone to subjectivity, prompting interest in the "LLM-as-a-Judge" paradigm~\cite{ZhengC00WZL0LXZ23}.

\citet{FaggioliDCDHHKK23} are among the first to investigate a range of human–machine collaboration strategies in which LLMs assist in relevance judgment. Building upon this, automated evaluations using LLMs have combined various prompting techniques such as zero-shot, one-shot~\cite{MacAvaneyS23}, or few-shot learning~\cite{TomasBing,upadhya1,upadhya2}. 

Another line of work~\cite{MaWYWL24,Fitte} involves training dedicated LLMs for assessment tasks.
Both \citet{MaWYWL24} and \citet{Fitte} explore the use of smaller, fine-tune LLMs for relevance assessment in real-world scenarios. However, such fine-tuned judge models often function as task-specific classifiers, thereby inheriting certain limitations in generalizability and reasoning capacity~\cite{Huang-abs-2403-02839}.
More recently, models such as JudgeLRM~\cite{judgelrm} and Rank-R1~\cite{rank-r1} have emerged, explicitly incorporating reasoning across different assessment tasks through RL with outcome-driven rewards. 

It is worth noting that some recent approaches have also integrated user behavior signals into the model input or reward to tackle personalized relevance. For example, \citet{chen2024towards} proposed the ProPBP that feeds user interaction data (clicks, dwell time, etc.) into the LLM judge to align its decisions with individual user preferences. However, such behavior signals are largely unavailable in the RAG system, rendering these methods inapplicable.

\section{UMbrela Prompt} \label{appendix:umbrela_prompt}
The following presents the full prompt used in UMbrela method, which is also employed by DeepSeek-R1 to generate reasoning chains and answers on unlabeled data.
\begin{tcolorbox}[title=Prompt]
\small
\textbf{System} \\
\textit{You are a relevance assessor working on a user-generated content platform.}\\
\textbf{User} \\
Given a query and a document, you must provide a score on an integer scale of 0 to 2 with the following meanings:\\
0 = represent that the document has nothing to do with the query \\
1 = represents that the document has some answer for the query, but the answer may be a bit unclear, or hidden amongst extraneous information \\
2 = represents that the document is dedicated to the query and contains the exact answer

\textit{Important Instruction:} \\
Assign category 1 if document presents something very important related to the entire topic but also has some extra information and category 2 if the document only and entirely refers to the topic. If none of the above satisfies give it category 0.\\
Please determine the primary intent behind a user's search query, using both your internal knowledge and the provided context.\\

Your response must strictly follow the format:\\
\texttt{<think> [the reasoning content] </think>} \\
\texttt{<score> [0/1/2] </score>}\\
\textit{Input}\\
\quad [query]: $\{query\}$ \\
\quad [document to be evaluated]: $\{doc\}$ \\
\textbf{Assistant} 
\end{tcolorbox}

\section{Instruction Template for R³A} 
\subsection{Prompt in the First Stage}\label{appendix:system_prompt_1}
The following presents the full prompt used in the first stage interaction.

\begin{tcolorbox}[title=Prompt in the 1st Stage]
\small
\textbf{System} \\
\textit{You are a content understanding engineer working on a user-generated content platform.}\\
\textbf{User} \\
Please determine the primary intent behind a user's search query, using both your internal knowledge and the provided context.\\
Your input consists of the [query] and the [in-platform documents] retrieved based on that query. The latter is intended to assist in judging the user's intent but may contain irrelevant content. The search query should be considered the primary reference.
Please carefully analyze the given [query] and the corresponding [in-platform documents] to infer the underlying query intent.\\

Your response must strictly follow the format:\\
\texttt{<think> [the reasoning content] </think>} \\
\texttt{<intent> [inferred user intent] </intent>}\\
\textit{Input}\\
\quad [query]: $\{query\}$ \\
\quad [in-platform documents]: $\{docs\}$ \\
\textbf{Assistant} 
\end{tcolorbox}

\subsection{Prompt in the Second Stage}\label{appendix:system_prompt_2}
The following presents the full prompt used in the second stage interaction.
\begin{tcolorbox}[title=Prompt in the 2nd Stage]
\small
\textbf{User} \\
Please assess the relevance of the [document to be evaluated] based on the user's input [query] and the inferred [intent], and extract the relevant fragment of the document accordingly.\\
\textit{Scoring Criteria}\\
\setlength{\parindent}{0em}
\quad 0 = not relevant, the document has nothing to do with the query. \\
\quad 1 = partially relevant, the document is relevant to the query but partly answers it.  \\
\quad 2 = highly relevant, the document is dedicated to the query and contains the exact answer.\\
\textit{Extraction Guidelines} \\
\quad 1. Extract the content from the [document to be evaluated] that is strictly relevant to the query and can help answer the query. This may include paragraphs, sentences, or even individual phrases.\\
\quad 2. The extracted content must come directly from the original document, with all punctuation preserved.\\

Your response must strictly follow the format:\\
\texttt{<think> [the reasoning content] </think>} \\
\texttt{<extract> [fragment/none] </extract>}\\
\texttt{<score> [0/1/2] </score>}\\
\textit{Input} \\
\quad [document to be evaluated]: $\{doc\}$ \\
\textbf{Assistant} 
\end{tcolorbox}

\section{Dataset Description} \label{appendix:dataset}
We collect a total of 50k unlabeled documents for cold-start training and distillation, and another 7k samples for human double-blind annotation. Our dataset has three classes: 
\begin{itemize}
    \item \textbf{0-Irrelevant}: There is a complete mismatch between the content of the query and the document.
    \item \textbf{1-Partially Relevant}: The document is relevant to the query and partly satisfies user's information needs.
    \item \textbf{2-Highly Relevant}: The document content is customized to satisfy the information needs of the query and precisely contains the answer to the query.
\end{itemize}

\subsection{Data Strategy}
\paragraph{Cold-Start (50k).} Following the data collection strategy detailed in Sec~\ref{sec:datasets}, we utilized document citation signals as a preliminary filter to curate an approximately balanced set of 50k unlabeled query-document pairs. Subsequently, we employed DeepSeek-R1 to generate structured CoT reasoning outputs. This phase is designed to finstill structured reasoning behavior and the desired output format.

\paragraph{RL Training (NoteRel 6k).} We utilize the NoteRel training set, comprising approximately 6k human-annotated samples as ground truth, to fine-tune all trainable models.

\paragraph{Distillation (50k).} We employ the same 50k unlabeled query-document pairs from the cold-start phase. These samples are labeled by the R³A-7B model, providing both reasoning traces and final scores. The query-document pair inputs with only score outputs are then used to distill knowledge into a compact 1.5B student model.

\subsection{Human Annotation}
To ensure the reliability of the annotations, we employed rigorous quality control measures, including:
\begin{enumerate}
    \item Annotation Guidelines: Annotators were provided with detailed guidelines and training sessions to standardize their understanding of relevance criteria, minimizing potential biases in the annotations. 
    \item Inter-annotator Agreement: A subset of annotations was double-annotated by two independent annotators to assess the consistency of the judgments. Only those annotations with an agreement above a predefined threshold were retained for further use.
    \item Feedback Loops: Regular feedback was provided to annotators to ensure continuous improvement of annotation quality during the process.
\end{enumerate}
These measures were implemented to ensure that the annotations are reliable, consistent, and representative of the true relevance of query-document pairs.

\section{Implementation Details} \label{appendix:training}
We employ Qwen2.5-1.5B(-Instruct) and Qwen2.5-7B(-Instruct) as the initial models. Models with the "-Zero" suffix are trained without cold start from the Instruct model. We utilize the OpenRLHF~\cite{openrlhf} framework for training. GRPO~\cite{grpo} is used as the RL algorithm.  We use the NoteRel as the training and test sets. We set the number of rollouts as 16 for one task. We set the learning rate as 5e-7, batch size as 32, training steps as 360. We set $\lambda$ as 0 in reward function. We use 8 A100 GPUs for all the experiments.




\section{Training Log of R³A-7B/1.5B}\label{appendix:log_fig}
\begin{figure}[t] 
    \centering
    \begin{subfigure}{0.43\textwidth}
        \includegraphics[width=\textwidth]{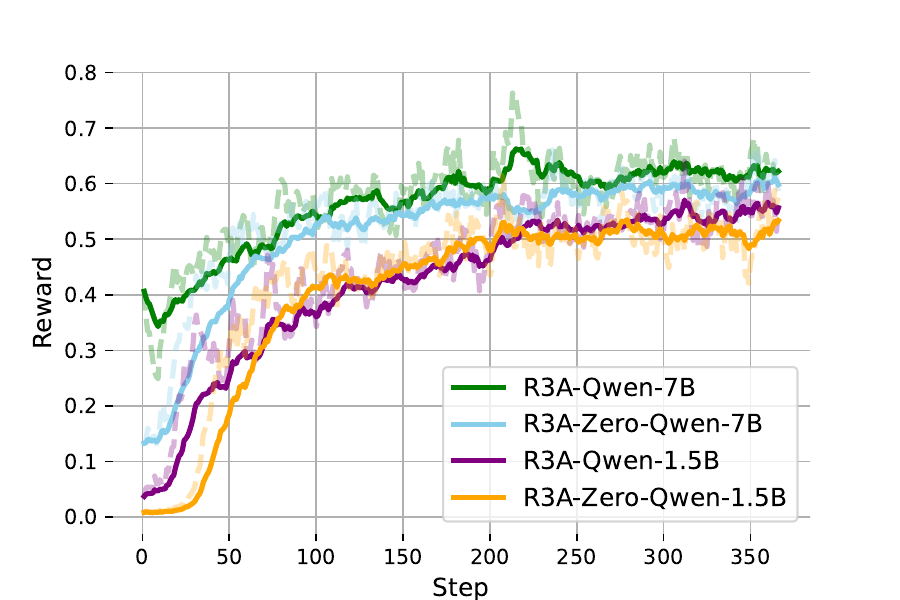}
        \caption{Training Rewards}
        \label{fig:reward}
    \end{subfigure}
\hfill
    \begin{subfigure}{0.43\textwidth}
        \includegraphics[width=\textwidth]{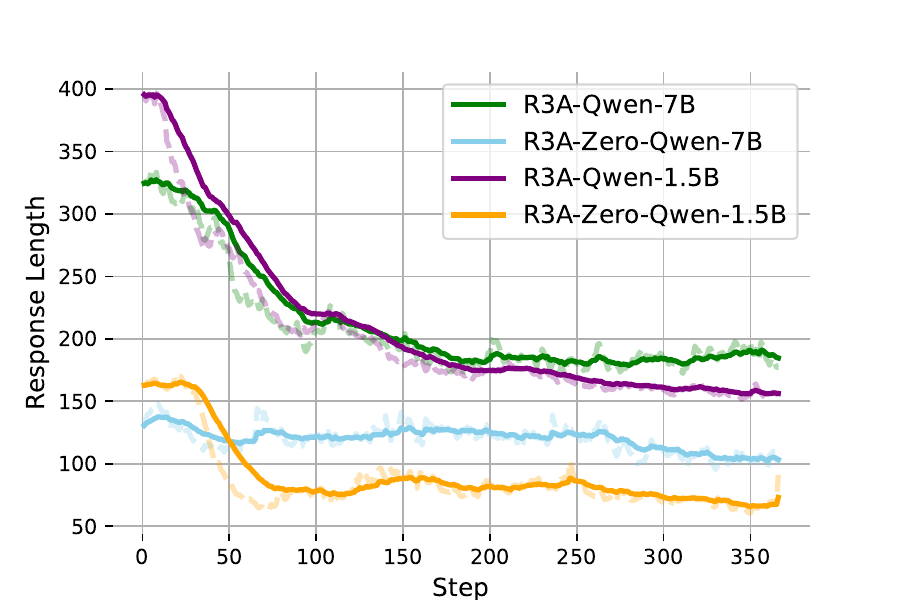}
        \caption{Response Length}
        \label{fig:length}
    \end{subfigure}
    \caption{The training log of R³A-Zero-1.5/7B and R³A-1.5/7B, including the curve of training rewards and response length.}
    \label{fig:line}
\end{figure}
Figure~\ref{fig:line} shows that R³A models initialized with cold start exhibit faster reward growth and achieve higher final rewards compared to their R³A-Zero counterparts, underscoring the benefit of cold-start initialization. Although Zero models begin with lower rewards, they gradually acquire the desired format, demonstrating that decomposed reasoning remains effective even without prior initialization. In terms of response length, both R³A and R³A-Zero models rapidly converge to a stable and concise output length.

\section{Case Study} \label{appendix:case_study}
Representative outputs from the $R^3A$ framework are provided in Tables~\ref{tab:icase} and \ref{tab:enhanced_case_study}. For brevity, the auxiliary high-clicked documents used for intent analysis are omitted to more clearly highlight the interaction between the query and document.

Example 1 illustrates a document with a gold score of 0, which was misclassified as partly relevant by the R1 model but correctly classified by R³A. 
The user query asked for the precise definition of an infinite series, whereas the document only contained related concepts from a table of contents and did not address the question directly. The R1 model was misled by superficial lexical matches and assigned a relevance score of 1, despite the document failing to meet the user's information need. In contrast, our model, R³A, successfully inferred the document's lack of relevance by identifying and reasoning over the actual content, ultimately producing the correct assessment. This case highlights a key strength of R³A: its ability to go beyond lexical similarity by integrating retrieved documents with deeper reasoning about content sufficiency, thereby producing more reliable relevance judgments in alignment with user intent.

In Example 2, the user inquires about the specific reasons for the absence of kindergarten tuition fees in Suzhou. The source document exhibits typical UGC noise, containing extraneous information such as personal anecdotes and traffic conditions. R³A demonstrates superior query disambiguation through its \textsc{<intent>} analysis, correctly identifying that the user's core intent is to verify the implementation details of a regional fee-waiver policy. By leveraging its decomposed reasoning capability, the model successfully isolates the pertinent policy evidence from the surrounding noise, achieving precise evidence localization (Score 2).

Example 3 highlights R³A’s ability to handle partial informational alignment in complex scenarios. The query concerns general ticket-purchasing procedures for seniors, while the document provides instructions specific to a seasonal promotion in the first quarter of 2024. The \textsc{<think>} output analytically recognizes the asymmetric relationship between the general query and the time-constrained evidence. The model correctly identifies that the document provides valid procedural steps but is limited by its seasonal nature, leading to an accurate partial relevance score (Score 1). This demonstrates the framework's robustness in maintaining high reasoning precision even when evidence is bound by specific temporal or conditional constraints.
\begin{table*}[ht]
\centering
\small
\renewcommand{\arraystretch}{1.1}
\resizebox{\textwidth}{!}{%
\begin{tabular}{|p{15cm}|}
\hline
\textbf{Example 1:}

\textbf{Query:} What does infinite series mean? \\
\textbf{Document:} Science students at the University of Manchester, speed up! The most comprehensive calculus vocabulary list is here
Hi dear students studying at UoM~ I’m sure your English is totally fine for daily communication, but when it comes to some technical terms, it can get tricky! Looking up unfamiliar words one by one is just too much trouble and really drags down our learning efficiency. When I was studying abroad, I often thought: if only there were a specialized vocabulary book for each subject!

Well, here it is! The most complete English-Chinese vocabulary list for calculus is finally here! You're really catching it at the right time!

\textbf{This vocabulary list is based on content from the following calculus topics}:

\textit{Limits and Continuity}: Definition and computation of limits; Infinite limits and asymptotes

\textit{Integrals}: 
Definitions of definite and indefinite integrals; Basic integration techniques: substitution, integration by parts

...

\colorbox{yellow!10}{\textbf{Infinite Series}}:

Convergence tests for series; Power series and Taylor series

\textit{Multivariable Calculus}: Partial derivatives and directional derivatives; Multiple integrals and vector calculus

To all students taking MATH10121, MATH10222, MATH20122, and MATH20411 at UoM — don’t let calculus vocabulary slow down your progress. Save this vocabulary list now, learn efficiently, and enjoy solving problems! Like + follow, and if there’s anything else you’d like to see, just DM me$~$
\#UniversityofManchester \#UoMstudyabroad \#UKstudents \#calculusQ\&A \#UoMstudents \\
\textbf{Gold Score: 0}  \\
\hline
\textbf{R1 Output:} \\
\texttt{<think>} First, I need to evaluate the relevance between the user's search term and the document to be assessed, and assign a score based on the evaluation criteria. The search term is "What does infinite series mean?", and the user's intent is to understand the concept and meaning of infinite series.

Next, let's look at the content of the document. The title mentions "The Most Comprehensive Calculus Vocabulary List" and covers multiple topics from calculus courses, including limits and continuity, derivatives, integrals, infinite series, and multivariable calculus. 

\colorbox{yellow!10}{\textbf{In the section on "Infinite Series," the document lists terms such as convergence tests and Taylor series.}}
When analyzing relevance, the core of the search term is the definition and meaning of "infinite series." Although the document does mention content related to infinite series, it primarily presents a vocabulary list of related terms and methods, such as convergence tests and power series, without providing a detailed explanation or mathematical definition of the concept. The purpose of the document is to offer a vocabulary list to help students memorize and learn technical terms in calculus, rather than to explain each concept in depth.

According to the scoring criteria, a score of 2 requires that the document directly meets the main need of the search term and provides a complete answer. Since the user is looking for an explanation of infinite series and the document does not provide detailed content in this regard—only related terms and methods—it partially addresses the topic. \textbf{Therefore, it should be scored as 1 point.}
\texttt{</think>} \\
\texttt{<answer>} \textcolor{red}{1} \texttt{</answer>} \\
\hline
\textbf{R³A Output in the 2nd Round (Ours):} \\
\texttt{<think>}
First, I need to assess the relevance between the user's search query and the document to be evaluated, and assign a score based on the evaluation criteria. The search term is "What does infinite series mean?", and the user’s need is to understand the definition or basic concept of infinite series.

Next, I examine the content of the document. Most of the document consists of a usage guide for a calculus vocabulary list and the source courses. In the "Infinite Series" section, it mentions "tests for convergence of series" and "power series and Taylor series." These may relate to certain evaluation methods and topics within infinite series, \colorbox{yellow!10}{\textbf{but they do not provide a definition or basic conceptual explanation of infinite series.}}

According to the scoring criteria, a score of 2 requires that the document directly satisfies the main need of the search term and provides a complete answer. The search term calls for a definition or basic concept of infinite series, which the document does not provide. A score of 1 is appropriate when the document partially answers the query or under specific conditions. In this case, \colorbox{yellow!10}{\textbf{the document does not provide a definition or basic concept of infinite series}}, but instead refers to some related methods and topics such as convergence tests and power series. Therefore, the document does not directly answer the search term and does not provide a complete response.

As a result, the document is irrelevant to the main intent of the search term and should be given a score of 0.
\texttt{</think>}\\

\texttt{<extract> None </extract>}\\

\texttt{<score>} \textcolor{green}{0} \texttt{</score>} \\
\hline
\end{tabular}
}
\caption{Example 1: The case comparison using the R1 and R³A models. The \colorbox{yellow!10}{highlighted content} indicate the portions in the original text that overlap with the query keywords, as well as the key reasoning components generated by the model.}
\label{tab:icase}
\end{table*}

\begin{table*}[t]
\centering
\small
\resizebox{\textwidth}{!}{%
\begin{tabular}{|p{15cm}|}
\hline
\textbf{Example 2}

\textbf{Query}: \textbf{Why haven't the tuition fees for Suzhou kindergartens been paid yet?} \\ 
\textbf{Document}: \colorbox{yellow!10}{\textbf{Suzhou kindergartens are all free!!!}} \textbf{It seems Suzhou is the only city in Jiangsu province with free kindergartens across the entire province.} The policy is great, saving us money! 
Every time we cross the street, buses stop far away to let us pass, so I have to drag my son along, afraid we'll be a step too slow. 

Jiangsu Province's free kindergarten policy will be gradually implemented starting from the fall semester of 2025.

The specific implementation time and scope are as follows:

2025: Kindergarten for the youngest children will be free.

2026: Kindergarten for the youngest and middle-aged children will be free.

2027: Kindergarten for the youngest, middle-aged, and oldest children will be completely free.

The implementation areas include:

Nanjing City: Pukou District, Qixia District, Yuhuatai District.

Wuxi City: Xishan District, Huishan District, Jiangyin City.

Xuzhou City: Xuzhou Economic and Technological Development Zone.

Changzhou City: Xinbei District, Liyang City.

Suzhou City: Gusu District, Wuzhong District, Xiangcheng District, Changshu City, Taicang City, Suzhou Industrial Economic and Technological Development Zone.

... \\ 
\textbf{Gold Score: 2}  \\
\hline
\textbf{R³A Output (two rounds)} \\
<intent> The user is inquiring about the reason for the absence of tuition fee collection in Suzhou. This likely involves policy changes or implementation delays. The intent is to confirm if the policy is now "free of charge" and understand the specific implementation details. </intent>

<think> The user wants to know why fees haven't been collected. The document directly addresses this by stating that kindergartens in Suzhou are now free and provides a detailed implementation timeline for different grades (2025-2027) and specific districts. Since the document directly satisfies the information need without requiring additional inference, it is highly relevant. </think>

<extract> \textbf{Suzhou kindergartens are all free!!! It seems Suzhou is the only city in Jiangsu with full-region fee waivers.} </extract> 

<score> \textcolor{green}{2} </score> \\ 
\hline
\textbf{Example 3}

\textbf{Query}: \textbf{How to buy Huangshan tickets for those over 60?} \\ 
\textbf{Document}: Must-read for coming to Huangshan, get the free and half-price activities in one post!! 

January Huangshan Scenic Area Ticket Pric\textbf{e}: 

Winter Tour Ticket Price: Winter Tour: Full ticket 150 yuan each, half ticket 75 yuan each. 

Winter Tour Ropeway Ticket Price: Winter Tour: Yungu and Taiping Ropeway tickets, 65 yuan each way; Yuping Ropeway ticket, 75 yuan each way; Xihai Grand Canyon Sightseeing Cable Car ticket, 80 yuan each way. 

January Huangshan Scenic Area 4 Major Privilege Activities: 

Activity: Free Ticket Opening Day Activity.

Activity time: Every Wednesday in the first quarter of 2024 (non-holidays), 12 days in total. 

Reservation process: 1. Tourists enter the "Huangshan Tourism Official Platform" Alipay or WeChat mini-program, click the "Free Ticket Opening Day" activity carousel on the homepage

... 

Activity: Free admission for youths under 18 and people aged 60 and above.

Activity content: \textbf{In the first quarter of 2024, youths under 18 and people aged 60 and above can enjoy free admission to Huangshan Scenic Area. Enter the Huangshan Tourism Official Platform mini-program Huangshan Scenic Area ticket reservation page, select the ticket type "Huangshan Ticket Free Consumption Voucher for Groups over 60" for reservation.} \\ 
\textbf{Gold Score: 1}  \\
\hline
\textbf{R³A Output (two rounds)} \\
<intent> Purchasing methods or steps for Huangshan scenic area tickets specifically for people over 60 years old. </intent>

<think> The user's need is the purchasing methods or steps for Huangshan tickets for people over 60 years old.The document provides the reservation process, including how to reserve through the Huangshan Tourism Official Platform mini-program and select the corresponding ticket type for reservation. Although the document primarily focuses on the free ticket activity, it has provided steps and methods for obtaining tickets, even if it is a free ticket case. However, the search term may also include purchasing tickets in non-free ticket situations (e.g., standard price purchase), but there is no other relevant information in the document. Therefore, \textbf{the document partially answers the question of how to buy tickets but may not cover all possible situations or detailed steps outside the promotional period, so the score should be 1}. </think>

<extract> In the first quarter of 2024, youths under 18 and people aged 60 and above can enjoy free admission to Huangshan Scenic Area. Enter the Huangshan Tourism Official Platform mini-program Huangshan Scenic Area ticket reservation page, select the ticket type "Huangshan Ticket Free Consumption Voucher for Groups over 60" for reservation. </extract> 

<score> \textcolor{green}{1} </score> \\ \hline
\end{tabular}
}
\caption{Example 2\&3: Case study of R³A's performance on test set of NoteRel dataset.}
\label{tab:enhanced_case_study}
\end{table*}

\section{LLM Usage Disclosure}
We use LLM for paper writing to check grammar and boost the clarity.  We do not use LLM to generate experiment code and analysis. 
\end{document}